# Reinforcement Learning based QoS/QoE-aware Service Function Chaining in Software-Driven 5G Slices

Xi Chen[1,2] | Zonghang Li[2] | Yupeng Zhang[2] | Ruiming Long[2] | Hongfang Yu*[2] | Xiaojiang Du (Senior Member, IEEE)[3] | Mohsen Guizani (Fellow, IEEE)[4]

[1] School of Computer Science and Technology, Southwest Minzu University, Chengdu, Sichuan, China
[2] Center for Cyber Security, UESTC, Chengdu, Sichuan, China
[3] Department of Computer and Information Sciences, Temple University, Philadelphia, PA, USA
[4] Department of Electrical and Computer Engineering, University of Idaho, Moscow, Idaho, USA

**Summary**

With the ever growing diversity of devices and applications that will be connected to 5G networks, flexible and agile service orchestration with acknowledged QoE that satisfies end-user's functional and QoS requirements is necessary. SDN (Software-Defined Networking) and NFV (Network Function Virtualization) are considered key enabling technologies for 5G core networks. In this regard, this paper proposes a reinforcement learning based QoS/QoE-aware Service Function Chaining (SFC) in SDN/NFV-enabled 5G slices. First, it implements a lightweight QoS information collector based on LLDP, which works in a piggyback fashion on the southbound interface of the SDN controller, to enable QoS-awareness. Then, a DQN (Deep Q Network) based agent framework is designed to support SFC in the context of NFV. The agent takes into account the QoE and QoS as key aspects to formulate the reward so that it is expected to maximize QoE while respecting QoS constraints. The experiment results show that this framework exhibits good performance in QoE provisioning and QoS requirements maintenance for SFC in dynamic network environments.

## 1 | INTRODUCTION

Communication networks have evolved through four major generations from 1G to 4G, which respectively features analog voice service, digitalized voice service, data service, and mobile broadband service. With the increment of diversity of devices and applications connected to 4G networks, network operators are faced with CAPEX (capital expenditure) and OPEX (operational expenditure) pressures due to the fact that revenues do not come proportionally to the massive investment, given the current technological architecture. Recent years have witnessed a large amount of efforts by different nations, companies, standardization bodied, etc., poured into the research and development of the 5th generation of communication networks, i.e., 5G networks. It is well acknowledged that 5G covers a wide spectrum of research topics in wired core networks as well as wireless networks[1], where network heterogeneity[2,3], security[4,5,6,7], mobility[8,9], etc., should be collectively considered to push forward its development.

Different entities hold different technological visions of 5G networks. According to reference[10], the architecture of 5G networks can be roughly divided as three layers, i.e., physical infrastructure layer, virtualized layer, service layer. The physical infrastructure layer holds various compute, storage, and network resources, which are abstracted as virtual resource pools to enable easy utilization and resource sharing. By invoking API exposed by virtualized layer, service layer delivers services oriented to end-users or devices.

The physical infrastructure layer is comparatively static while the virtualized layer on top of that is dynamic. More concretely, it can be assumed that the number and configurations of physical compute, storage, and network resources are comparatively immutable in the short term, while the number and configurations of VMs (Virtual Machines) and VNF (Virtualized Network Functions) instances are mutable on demand to support their application-specific service layers. This implies two entailments. On one hand, the underlying network resources should be "sliced" for different service renderers (i.e., tenants) with minimal conflicts, i.e., the concept of *slicing* where network is vertically tailored into multi-layer *slices* independently controlled and managed by corresponding tenants.

On the other hand, network services nowadays are orchestrated by different network functions (NF, e.g., firewall, deep packet inspection, WAN optimizer, proxy, etc.), often in a virtualized fashion (i.e., VNF) to provide required functionalities as well as possibly improving QoE and maintaining QoS. This is where *service function chaining* (SFC)[11] comes into the picture. Therefore, an SFC framework that is QoS/QoE-aware inside a slice is a key in successful service orchestration and delivery in 5G core networks. While MEC (Mobile Edge Computing)[12] aims to handle issues related to 5G edge networks, SDN (Software-Defined Networking)[13] and NFV (Network Function Virtualization)[14] are considered key enabling technologies for 5G core networks which fabricate the backbone of 5G ecosystems and offer critical network services. We envision that at least three aspects should be addressed for SFC orchestration with regard to the highly dynamic SDN/NFV-enabled 5G slices.

1. A "smart" orchestration agent that is adaptive to the changing environment so that it can learn to approximate the optimal SFC orchestration policy with minimal human interference for automation purpose, i.e., the learning aspect.

2. A lightweight mechanism to evaluate the QoE of a service function chain in changing environments so that the agent can learn to maximize the user experience which is the key factor for 5G user subscriptions, i.e., the awareness aspect.

3. The ability to explore VNF alternatives that can potentially orchestrate the chain (e.g., for the purpose of load-balancing), while exploiting the best known VNF instances to optimize QoE, i.e. the exploration-exploitation aspect.

Reinforcement learning[15], with its trial-and-error mechanism (for aspect 1), reward mechanism (for aspect 2), exploration-exploitation ability (for aspect 3), etc., makes a competitive candidate for the SFC orchestration framework in 5G slices. Meanwhile, recent years have also seen its application in modern network paradigms for user experience optimization[16], cost minimization in resource allocation[17], etc. Based on the discussion above, this paper proposes the reinforcement learning based QoS/QoE-aware service function chaining framework in the context of software-driven (i.e., SDN/NFV-enabled) 5G slices. The authors believe the work presented by this paper addresses some missing parts of the current research on SFC. On one hand, many previous works abstract the SFC problem as a mathematical programming problem and present heuristic algorithms to balance efficiency and optimality. The quality of the service function chain is usually judged by the derived cost (the smaller the better) or utility (the bigger the better). The calculation of cost/utility usually depends on QoS metrics (such as bandwidth, delay, throughput, etc.). Existing literature often assumes that these metrics have been readily collected through some mechanism behind the hood. However, the collection of QoS information of various entities is often challenging and needs an elaborate design to implement a lightweight framework. On the other hand, although QoS is well studied with regard to SFC, QoE is not extensively considered in previous works. QoE and QoS exhibits a non-linear mutual relationship; thus the guarantee of QoS dose not necessarily ensure highly acknowledged QoE. Therefore, QoE should be explicitly considered for SFC. To summarize, the contribution of this paper is twofold:

- A lightweight QoS information collecting scheme with regard to SDN deployment in 5G slice. This scheme uses LLDP[18] as the "ferry" to load QoS information collected from underlying switches in a piggyback fashion so that no fundamental modification of the current OpenFlow-based southbound interface needs to be made. QoS information collecting is helpful in evaluating QoE and maintaining QoS constraints.

- A MDP (Markov Decision Processes) modeled reinforcement learning based service function chaining algorithm is proposed in the context of NFV. This algorithm takes into account the QoE and QoS as key aspects to formulate reward so that it is expected to maximize QoE while respecting QoS constraints.

The remainder of this paper is organized as follows. Section 2 summarizes related works and makes a brief comparison to our work. Section 3 exhibits the general system architecture. Section 4 explains the novel lightweight QoS information collecting scheme and how the collected information is used to simplify the network topology. Section 5 gives the details of MDP-modeled reinforcement learning based QoS/QoE-aware SFC framework. Section 6 takes experiments on both the novel lightweight QoS collecting scheme and the QoS/QoE-aware service function chaining. Finally, this paper is concluded in Section 7.

## 2 | RELATED WORKS

### 2.1 | Service Function Chaining

The trend of software-ized control led by SDN and virtualized network functions led by NFV has made the flexible service function chaining feasible which is also knowns as service/middlebox chaining. StEERING[19] extended OpenFlow and the NOX controller and implemented middlebox-based service chaining. The key of the extension is the split of a monolithic flow table into several micro tables to constrain the "rule explosion (i.e., too many rules)" during the mapping between service function chains and rule table entries. StEERING abstracts the service chaining as a graph theory problem for which a greedy algorithm combined with heuristics was proposed to solve it.

SIMPLE[20] proposed to conduct service function composition in the context of SDN. Service composition is split into 2 stages: online stage and offline stage. During the offline stage, the TCAM capacity is treated as the primary constraint based on which Integer Linear Programming is adopted to solve the problem; during the subsequent online stage, a simplified Linear Programming algorithm is used to tackle the load-balance problem.

FlowTags[21,22] holds the opinion that it is difficult to track the states of user traffic while traversing operator networks, which might result in the incorrect enforcement of network-wide policies. It is, therefore, also difficult to construct a service function chain to satisfy a user's business logic requirements and policy requirements of operator networks. FlowTags tags the traffic traversing middleboxes in a way that context information of middleboxes are organized as packet header tags shared along the service function chain. It can be invoked through southbound interface APIs so that the correctness of the service function chains and the consistency of network-wide policies are both possibly guaranteed.

References[23,24] advocated to use the named NF instances to facilitate the decoupling of the service plane and the data plane, so that the execution of service instances need not locate concrete positions of these instances, such as IP addresses, etc. Meanwhile, traffic steering is conducted according to the instance names stored in the packet headers, without being translated to flow table entries stored in switches, so that the size of the flow table is contained. On the contrary, switches store only the mapping of instance names and their IP addresses.

Reference[25] studied the function composition that maximizes throughput. Two algorithms (namely TMA and PDA) are designed corresponding to offline requests (i.e., those requests whose traffic characteristics can be determined by historical data or SLA (Service Level Agreement)) and online requests (i.e., those requests whose traffic characteristics can only be determined upon arrival). Both algorithms try to solve the utility-based (i.e., the throughput) optimization problem.

Reference[26] studied the NF consolidation problem, attempting to deploy more NFs onto fewer physical nodes so that utilization of NFs in the entire network can be made as high as possible. Regarding problem modeling, it is modeled as an integer programming problem, which is solved using IBM CPLEX optimization software in small-scale networks. For large-scale networks, a greedy-based heuristic algorithm is designed to solve a configuration that tries to minimize the number of VNFs. Similar work was conducted by reference[27], whose focus is the service function chaining in data centers. By taking into account the resource (computing, bandwidth, etc.) consumption by computing and switching devices, the energy utilization model is established for data centers. Besides, the traffic intensity is used to express the affinity between service functions, based on which service functions with high affinities could be placed nearby or on the same physical server. In this way, extra energy consumption due to long distance interaction is minimized.

The works presented by references[19,20,21,22,23,24] are functional service function chaining with little QoS consideration, whereas those by references[25,26,27,28,29] consider QoS metrics or other relative properties while QoE is not considered. On the other hand, our work takes into account both QoS and QoE in service function chaining.

### 2.2 | MDP-based Path/Chain Composition

MDP has long been used in resource composition, such as Web services composition in early works[30]. It has also been extensively used in network path and service chain composition recently. Reference[31] proposes the QoS-Aware Routing based on reinforcement learning. It gives a reward model that takes into account the QoS metrics such as bandwidth, delay, etc., and a Softmax-based policy to choose the next-hop forwarding device. The methodology is similar to our work. However, it is applied in forwarding plane routing with QoS awareness. Also, it does not combine the QoS constraints with reinforcement learning framework like our work. Reference[17] proposes the MDP model for NFV resource allocation to form a service chain with cost optimization. It adopts the Bayesian probability to predict the transition probabilities among VNF instances, thus it boils down to the model-based reinforcement learning where transition probabilities are fully observable and the value iteration approach is used for solution. However, in their framework, QoS awareness is not deeply considered.

Our work differs from the previous works in that both QoS and QoE are considered in the reward modeling.

## 3 | SYSTEM ARCHITECTURE

The ETSI NFV specification is believed to be one of most suitable to deploy our QoS/QoE-aware SFC framework. In addition, we also advocate the deployment of an SDN controller inside of a 5G slice on the control plane to implement QoS information collecting along with the topology discovery. Based on ETSI specifications, reference[32] proposes a network slice management and orchestration (MANO) architecture for 5G networks, as shown in Figure 1 , where the VIM (Virtual Infrastructure Manager, e.g., OpenStack, KVM, etc.) corresponds to the Infrastructure Manager, the VNFM (Virtual Network Function Manager) corresponds to the Network Slice Manager, while the NFVO (Network Function Virtualization Orchestrator, e.g., OpenStack Tacker) corresponds to the Service Instance Layer. We extend this architecture by implementing a QoS information collector as an SDN controller module (the green rounded rectangle in the middle on the right. See details in Section 4.1). We also extend the NFV MANO (Management and Orchestration) with a reinforcement learning based QoE/QoS-aware SFC agent as a service manager module (the green rounded rectangle on the top of the right. See details in Section 5).

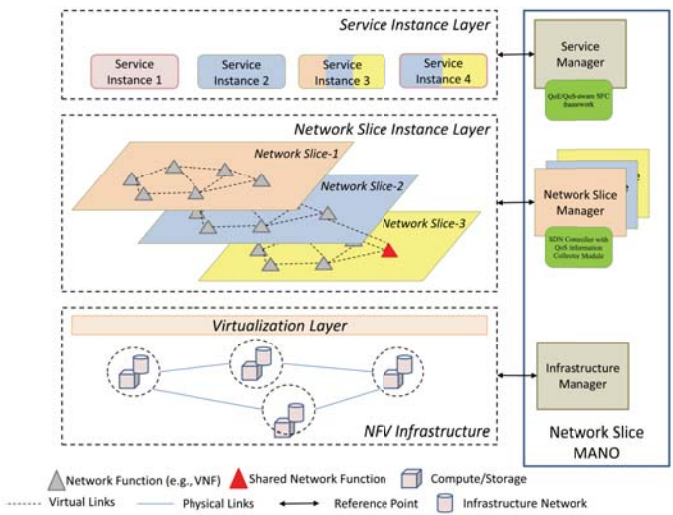

**FIGURE 1** Network Slice Management and Orchestration (MANO) Overview.

## 4 | QOS INFORMATION COLLECTING AND TOPOLOGY SIMPLIFICATION

### 4.1 | QoS over LLDP Scheme

In standard SDN networks, controllers are aware of switches directly connected to them through bidirectional Hello messages in the standard OpenFlow protocol. However, the underlying link states between switches (i.e., how switches are mutually connected) are not visible to controllers in the first place. Therefore, in the initial stage of an SDN network, controllers do not have the topology knowledge of the whole SDN network. In order to perform centralized control over an SDN network, controllers must carry out topology discovery. Controllers usually use LLDP to fulfill such a task. LLDP is an IEEE proposed protocol widely used in network arena for topology discovery.

The controller instructs a switch to multicast the LLDP packet to all of its ports through a packet-out (instructive packets from controllers to switches). In this packet-out, topological information of the switch such as chassis information, port information, etc., is all contained. All other switches connected to this sender switch receive the LLDP packet, and then match this packet against the flow table entries of their own, only to find no matches for LLDP packets. Thus, switches will send a packet-in (packets from switches to controllers) containing this LLDP to the controller asking how to process this packet. Since the packet-in contains topology information about both the sender switch and the receiver switch, the controller can now assert that there exists a link between the two switches based on the received packet-in. By means of this iterative packet-in/out interaction, topology of the whole SDN network can be discovered by the central controller. This centralized topology discovery in SDN is quite different from how LLDP works in traditional networks where topology discovery is done by individual switches independently, although LLDP is used in both cases.

Standard LLDP packets usually contain basic information such as the MAC address, chassis information, port information, etc. We can see from the above topology discovery phase, no QoS information is contained in LLDP packets. Should QoS information be incorporated, the QoS-aware topology discovery can be done to enable further QoS-aware decisions and policies, thus QoS provisioning becomes possible.

LLDP is a TLV (Type/Length/Value, i.e., key-value pair with length information) based protocol where TLVs are used for property descriptions. We can include QoS information as custom TLVs in LLDP packets. In this way, LLDP can be seen as the "ferry" containing QoS information (i.e., QoS over LLDP) and other useful properties as its payload. We define QoS TLV as follows in Figure 2 .

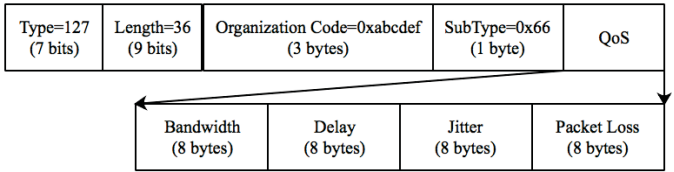

**FIGURE 2** QoS over LLDP Packet Format.

In the TLV Type field, it must be designated as 127 to indicate that this is a custom TLV. The Length field specifies the variable-length value contained in the TLV. The Organization Code field indicates the designer of this customized TLV. We use the Organization Code as 0xabcdef for the time being. The Subtype field specifies the detailed type of the contained value. The Value String field (i.e., the QoS field in Figure 2) gives the real value. We contain various QoS metrics in the Value String. In order for the receiver to conveniently parse the different QoS metrics, we use the predefined property order and length. We can see from Figure 2 that several metrics are included in fixed length in our current settings, namely delay, bandwidth, packet loss, and jitter, 8 bytes for each property. Therefore, a QoS over LLDP packet is 38 bytes longer than a pure LLDP packet in length. Note that more metrics such as availability can be included in the future work. Upon receiving the QoS over LLDP packet, the switch fills QoS metrics in corresponding TLV fields. We have implemented this mechanism in Floodlight controller and OVS (Open vSwitch)[33].

## 4.2 | Topology Simplification

To orchestrate a service function chain is to chain a set of VNF instances distributed on virtual machines or containers initiated on physical commodity servers, usually one instance per VM/container. Servers are inter-connected by physical/virtual forwarding devices (e.g., switches) and links. According to reference[34], the total number of middleboxes (i.e., VNF instances in the context of NFV) is comparable to the number of forwarding devices in modern ISP networks or datacenters. Therefore, if entities in the forwarding plane are explicitly involved during the process of SFC orchestration, which essentially boils down to a service plane problem, the problem complexity is much greater than that only VNF instances are taken into account. However, if forwarding entities are not explicitly considered during orchestration, the forwarding-plane datapath is still needed to be planned separately after orchestration, as a second step, for traffic steering that sequentially traverses VNF instances, resulting in a two-tier solution.

If forwarding devices and links are collectively viewed as an "aggregated link" between two servers hosting VNF instances, the topology can be simplified as one with VNF instances as nodes and "aggregated links" as edges, without the involvement of forwarding-plane entities. Therefore problem complexity can be reduced during orchestration. Note that QoS over LLDP collects QoS information alongside topology discovery, which means that the QoS status of aggregated links can be mathematically inferred to support quality evaluation of orchestration. Let $v_{ij}$ denote the real value of the $j$-th QoS metric of device $i$ in an aggregated link, where $j \in \{dl, bw, pl, av, jt\}$, namely *delay*, *bandwidth*, *packet loss*, *availability*, and *jitter*, the algorithms to evaluate the QoS metrics of an aggregated link are shown as follows in Table 1.

**TABLE 1** QoS of the Aggregated Link.

| QoS Metric | Aggregation Algorithm |
|---|---|
| dl (delay) | $\sum_{i=1}^{n} v_{ij}$ |
| bw (bandwidth) | $\min_{i \in n} v_{ij}$ |
| pl (packet loss) | $1 - \prod_{i=1}^{n}(1 - v_{ij})$ |
| av (availability) | $\prod_{i=1}^{n} v_{ij}$ |
| jt (jitter) | $\sum_{i=1}^{n} v_{ij}$ |

Now that forwarding-plane entities (e.g., switches, links, etc.) are consolidated as aggregated links between VNF instances, the topology can be, at large, simplified. Figure 3 gives a simple example of topology simplification. Server-1, which hosts a VNF instance v-ins-1, is connected to Server-2, which hosts v-ins-2, v-ins-3 and v-ins-4, through switch-1 and switch-2 via 3 links. QoS over LLDP discovers the forwarding topology with corresponding QoS metrics, shown in the bottom with solid squares. Switch-1, Switch-2 and links in between form an aggregated links whose QoS metrics are inferred according to Table 1. Therefore, v-ins-1 is connected to v-ins-2, 3 and 4 with an aggregated link with 25 us delay and 100 Mbps available bandwidth (i.e., the dashed square).

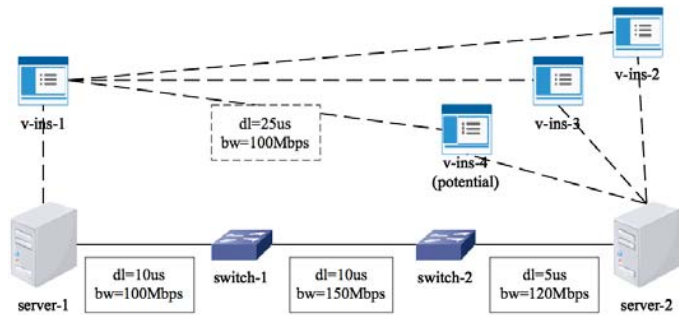

**FIGURE 3** An Aggregated Link Example.

# 5 | REINFORCEMENT LEARNING BASED QOS/QOE-AWARE SFC

## 5.1 | Problem Statement of SFC Orchestration

Suppose an SFC request imposes $N$ network functions (e.g., traffic sequentially passes through firewall, DPI, etc.). Each function can be accomplished by a VNF type, $t_i$, $i \in \{1, 2, \cdots, N\}$, and each VNF type $t_i$ has $M_i$ candidate VNF instances. Let $ins_{ij}$ denote the $j$-th VNF instance of VNF type $t_i$, $i \in \{1, 2, \cdots, N\}, j \in \{1, 2, \cdots, M_i\}$.

**Definition 1** (Functional SFC Orchestration, F-SFC). Let $x_{ij} \in \{0, 1\}$ denote whether $ins_{ij}$ is selected ($x_{ij} = 1$) or not ($x_{ij} = 0$) to accomplish the $i$-th function required by the SFC request. The functional SFC orechestration (F-SFC) is to select one and only one VNF instance from $t_i$ for the $i$-th function.

$$\sum_{j=1}^{M_i} x_{ij} = 1$$
$$\sum_{i=1}^{N} \sum_{j=1}^{M_i} x_{ij} = N \quad (1)$$

The previous definition implies that there should exist $M_i$ *deployed VNF instances* so that one of them can be "selected" to fulfill the $i$-th function. However, in the context of SDN/NFV, a VNF instance can be instantiated on-demand without prior existence. In other words, the remaining resources of the commodity server that hosts VNF instances can be seen as the *potential VNF instances* (e.g., v-ins-4 in Figure 3 ) as long as there are enough resources for instantiation. To capture this dynamic nature of VNF instantiation, we regard the remaining resources of the direct successive commodity server as a potential VNF instance for the algorithm to select from. This instantiate-then-select operation is different from pure selection from existing VNF instances in that it incurs extra booting time, extra power consumption, extra operational activities, etc., which can be considered as operational expenditures (OPEX). In this regard, equation (1) covers the both the deployed VNF instance selection and on-demand VNF instantiation commonly seen in SDN/NFV scenarios.

**Definition 2** (QoE-aware SFC Orchestration, QoE-SFC). Let $C$ denote the set of all service function chains that can functionally satisfy the SFC request. Let $qoe_c$ denote the the end-to-end QoE of service function chain $c \in C$. QoE-aware SFC orchestration (QoE-SFC) is the F-SFC that maximizes the end-to-end QoE of all candidate chains. The QoE evaluation of $qoe_c$ will be discussed in Section 5.3.1.

$$\max_{c \in C} qoe_c$$
$$s.t. \quad \sum_{j=1}^{M_i} x_{ij} = 1$$
$$\sum_{i=1}^{N} \sum_{j=1}^{M_i} x_{ij} = N \quad (2)$$

**Definition 3** (QoE/QoS-aware SFC Orchestration, Q2-SFC). Let $qos_c$ denote an $L$ dimensional vector that indicates the QoS metrics of service function chain $c \in C$. Let $qcon$ denote an $L$ dimensional vector that indicates the QoS constraints of the SFC request. Without losing generality, let assume that the first $K$ dimensions of the QoS vector that are positive metrics (i.e., the greater values the better, e.g., bandwidth) and the $(L - K)$ remaining dimensions of the QoS vector are negative metrics (i.e., the smaller values the better, e.g., delay). QoE/QoS-aware SFC Orchestration (Q2-SFC) is the QoE-SFC that satisfies the QoS constraints of the SFC request.

$$\max_{c \in C} qoe_c$$
$$s.t. \quad qos_c^t \geq qcon^t, t \in \{1, 2, \cdots, K\}$$
$$qos_c^t \leq qcon^t, t \in \{K+1, K+2, \cdots, L\}$$
$$\sum_{j=1}^{M_i} x_{ij} = 1$$
$$\sum_{i=1}^{N} \sum_{j=1}^{M_i} x_{ij} = N \quad (3)$$

Note that the topology to be dealt with is a simplified topology using aggregated links to reduce complexity. Also the QoS vector $qos_c$ can be formulated by the QoS information collected by QoS over LLDP scheme.

During the SFC orchestration process, two strategies can be adopted:

- Incremental orchestration: The selection of VNF instances for functions are conducted in a hop-by-hop fashion. Therefore, the length of the service function chain gradually increases.

- Monolithic orchestration: Every step gives a complete service function chain (i.e., one VNF instance is selected for each function) and checks whether it maximizes QoE and meet QoS constraints. If not, find another complete service function chain in the next step.

In this paper, we adopt the incremental orchestration strategy due to that 1) it can be easily mapped to a multi-step reinforcement learning problem modeled using MDP; and 2) it is finer-grained, and thus a sophisticated policy can be derived for QoE maximization and QoS constraints. Meanwhile, we envision that the monolithic strategy fits better in QoS maintenance during the runtime of an existing chain, which is out of the scope of this paper.

Our goal in this paper is to implement Q2-SFC using reinforcement learning. The reasons why we choose reinforcement learning for Q2-SFC solution lie as follows:

- Fewer requirements are needed during training by reinforcement learning, compared with supervised learning, in that no prior extensive training dataset is required. Training dataset from real-world networks is hard to acquire. On the one hand, great efforts are required both in computing and storage to store operational statistics. On the other hand, a dataset might (unintentionally) contain or infer sensitive data which is why network operators are not quite willing to share.

- The model trained by supervised learning can hardly reflect the dynamics of a continuously changing network environment. On the contrary, through the reward mechanism, reinforcement learning can better adapt to environmental changes.

## 5.2 | The MDP of SFC Orchestration

The MDP usually consists of five ingredients, i.e., $\{S, A, P, R, \gamma\}$ where $S$ denotes the finite set of *states* [1], $A$ denotes the finite set of *actions*, $P$ denotes the finite set of *state transition probabilities*, and $R$ denotes the finite set of *immediate rewards*. $\gamma \in [0, 1]$ is the discount factor, indicating the importance of future of rewards to the current reward. The solution of the MDP is called a *policy* given that in the current state $s \in S$, an action $a \in A$ is selected to maximize the long term rewards. If the state transition probability $p^a_{s \to s'}$ from state $s$ to $s'$ given that action $a$ is selected is unknown, the model of the MDP is unknown. In that case, the solving of the MDP is called model-free reinforcement learning.

Note that the quality of a policy $\pi(s, a)$ is not determined by the immediate reward $r$; instead, it is evaluated by long term rewards, thus two value functions are defined to capture this: the state value function $V^\pi(s)$, which indicates the expected accumulated discounted rewards from initial state $s$; and the action value function $Q^\pi(s, a)$ (also called state-action value function), which indicates the expected accumulated discounted rewards by action $a$ from initial state $s$. Mathematically, we have the following:

$$V^\pi(s) = E(\sum_{k=0}^{\infty} \gamma^k r_{t+k+1} | s_t = s) \quad (4)$$
$$= E(r_{t+1} + \gamma V^\pi(s_{t+1}) | s_t = s)$$

$$Q^\pi(s, a) = E(\sum_{k=0}^{\infty} \gamma^k r_{t+k+1} | s_t = s, a_t = a) \quad (5)$$
$$= E(r_{t+1} + \gamma Q^\pi(s_{t+1}, a_{t+1}) | s_t = s, a_t = a)$$

where $r_t$ indicates the immediate reward of step $t \in \{1, 2, \cdots, T\}$; $E(\cdot)$ is the mathematical expectation operator. To find the optimal solution of MDP is to find the policy that maximizes state value function:

$$\pi^* = \arg\max V^\pi(s) \quad (6)$$

According to Bellman Optimality Equation, the optimal policy $\pi^*$ is the one that holds the following:

$$V^{\pi^*}(s) = \max_{a \in A} Q^{\pi^*}(s, a) \quad (7)$$

We model the Q2-SFC orchestration as an MDP in that:

- $S$: Every state $s \in S$ represents the system environment including network topology, VNF instances' QoS/QoE status, functional and QoS requirements of the SFC request being processed, etc.

- $A$: Every action $a \in A$ represents the selection of a certain direct successive VNF instance from the current VNF instance. Obviously, for the $i$-th function there exist $M_i$ actions (selections).

- $P$: Every transition probability $p^a_{s \to s'} \in P$ represents the possibility that the QoS/QoE status changes from $s$ to $s'$ under VNF instance selection action $a$. However, it is unknown here thus a model-free reinforcement learning.

- $R$: Every immediate reward $r \in R$ represents the contribution of the selected VNF instance $ins_{ij}$ to the current QoE of the chain.

The solution of MDP-modeled Q2-SFC is to find the optimal service function chain $c^* \in C$ (where $C$ is a finite set of service function chains) under policy $\pi$ such that:

$$c^* = \arg\max_{c \in C} E(\sum_{t=0}^{T} \gamma^t r_{t+1}) \quad (8)$$

## 5.3 | The Reward Design

According to equation (8), the key to solve MDP-modeled Q2-SFC is the reward model that reflects QoE, possibly under QoS and other constraints, and the policy design that maximizes long term rewards. Therefore, the reward design should cover aspects not just QoE. We consider the QoE gain, the QoS constraints penalty, and the OPEX penalty to be ingredients that constitute the overall reward of an action during incremental SFC orchestration.

### 5.3.1 | The QoE Gain

The evaluation of QoE can be roughly divided into two categories, namely subjective and objective evaluation. The subjective evaluation involves end-user's participation in rating the service from the perspective of direct user perception. Usually, the MOS (Mean Opinion Score)[35] scale is used during subjective evaluation. Although it has advantages, such as intuitiveness, accuracy, etc., in evaluating service experience, subjective evaluation requires great efforts in mobilizing user participation and is subjected to the varying understanding

---
[1] We can see that infinite states can also be dealt with in reinforcement learning in later sections.

**8** | XI CHEN ET AL
and preferences of the various experience metrics. Therefore, subjective evaluation can hardly be applied in large networks, whereas its main application lies within the evaluation of other QoE evaluation methods, such as objective evaluation.

Objective evaluation, on the contrary, derives QoE from measurable metrics without end-user involvements, thus the automation of QoE evaluation becomes possible. QoS metrics are prominently used in the automated QoE evaluation, among others, where QoE is calculated as per measured QoS metrics as well as the consideration of psychological perception from end-users. Two well-known principles, i.e., the WFL (Weber-Fechner Law)[36] and IQX (Exponential Interdependency of QoE and QoS) hypothesis[37], are used in the deriving from QoS to QoE, both of which give non-linear relationship between QoS and QoE as shown in the following equations:

$$dQoS \propto QoS \cdot dQoE, \text{WFL} \quad (9)$$

$$dQoE' \propto QoE' \cdot dQoS, \text{IQX} \quad (10)$$

Although, seemingly, WFL and IQX give contradictory relationships between QoE and QoS (i.e., differential v.s. exponential), we argue that WFL and IQX apply in QoS metrics with different tendencies. For positive QoS metrics (the bigger value the better), the corresponding QoE can be derived in equation (11) according to WFL while for negative QoS metrics (the smaller value the better), the corresponding QoE can be derived in equation (12) according to IQX, where $qos_c^t$ is the $t$-th QoS metric of service function chain $c$, whose value can be calculated using algorithms in Table 1 . $\alpha_p, \beta_p, \gamma_p, \theta_p, \alpha_n, \beta_n, \gamma_n$ and $\theta_n$ are constant parameters to fine-tune QoS/QoE relationships. Study and fine-tuning of these parameters are out of the scope of this paper. Interested readers can refer to reference [36,37] for detailed mathematical relationships.

$$qoe_c^t = \gamma_p \times \log(\alpha_p \times qos_c^t + \beta_p) + \theta_p, t \in \{1, 2, \cdots, K\} \quad (11)$$

$$qoe_c^t = \gamma_n \times e^{\alpha_n \times qos_c^t + \beta_n} + \theta_n, t \in \{K+1, K+2, \cdots, L\} \quad (12)$$

To give better intuitions of these equations, we give some daily experiences as examples. Suppose a user has a 10 Mbps access bandwidth (i.e., a positive QoS metric). According to daily experience, the increase of bandwidth to 20 Mbps does not give the user the perception that the speed is twice as fast; on the contrary, it gives very limited perception of speed upgrade. However, if the bandwidth is upgraded to 100 Mbps, the perception of speed upgrade is somehow obvious. This can be well captured by equation (11) (i.e., WFL). For another example, glitches or paused buffering in video streaming caused by minor packet loss (i.e., a negative QoS metric) would greatly compromises the user experience, resulting in their possible refreshing of the Web pages impatiently, which is captured by equation (12) (i.e., IQX). The overall QoE of a service function chain can be derived by the following equation:

$$qoe_c = \sum_{t=1}^{K} w^t \times qoe_c^t - \sum_{t=K+1}^{L} w^t \times qoe_c^t \quad (13)$$

The QoE gain of constructing a chain $c$ by taking action $a$ (i.e., selecting all those VNF instances $ins_{ij}$s that form chain $c$) under state $s$ is shown as follows in equation (14). Note that how to select VNF instances $ins_{ij}$s to construct a chain $c$ will be discussed in Section 5.5.

$$gain_c^{qoe} = qoe_c \quad (14)$$

### 5.3.2 | The QoS Constraints Penalty

Intuitively, QoE can be used as the estimate of immediate reward. In this way, service function chain with the highest accumulated rewards is considered the best one, which also maps well to the standard reinforcement learning whose solution is the answer to QoE-SFC that maximizes QoE. Nevertheless, in the standard reinforcement learning model, no constraints are explicitly specified. If we adopt the standard reinforcement learning in Q2-SFC and simply regard QoE as reward, no QoS constraints are enforced. Therefore, we believe that the standard reinforcement learning does not serve well in Q2-SFC. The key to adapt reinforcement learning to Q2-SFC is to embrace QoS constraints. However, if we explicitly specify QoS constraints in reinforcement learning as do in mathematical programming approaches (like that in Definition 3), we are very likely to face the NP-hardness that leads to an impractical solution within polynomial time.

Obviously, there exists a paradox between QoE and QoS, in that maximizing QoE requires high resource consumption, while respecting QoS constraints requires low resource consumption. High resource consumption narrows the "distance" between QoS metrics and QoS constraints. If the "distance" between the QoS metrics of the chain and the QoS constraints is very close, the probability of violating QoS constraints is high. This should generate a penalty against the reward, i.e., a negative reward. The closer the "distance", the bigger the penalty against the reward. If any QoS metric violates the corresponding constraint, the penalty is considered very severe. To capture this, we define the penalty due to the distance between QoS metrics $qos_c$ of chain $c$ and the QoS constraints $qcon$ as follows in equation (15), where $P$ is a large-enough constant to penalize QoS constraints violations.

$$pen_{ij}^{qcon} = \begin{cases} P, & \text{if any QoS constraint violation} \\ P \cdot e^{-\sqrt{\sum_{t=1}^{L} ||qos_c^t - qcon^t||^2}}, & \text{otherwise} \end{cases} \quad (15)$$

## 5.3.3 | The OPEX Penalty

In the discussion of Definition 1, we distinguished *deployed VNF instances* and *potential VNF instances* which captures the nature that VNF instances can be instantiated on-demand by sufficient remaining resources on physical commodity servers hosting VNF instances. The instantiation of new VNF instances shall suffer from great OPEX since it might incur the loading of remote virtual machine images stored in image repositories and the instantiation of VNF instances, which as well we consider a penalty against rewards. The operational expenditure for VNF instance $ins_{ij}$ is defined as follows:

$$pen_{ij}^{opex} = \begin{cases} opex^{normal} + opex_i^{vm} + opex_i^{vnf}, & \text{if } ins_{ij} \text{ is potential} \\ opex^{normal}, & \text{otherwise} \end{cases} \quad (16)$$

$opex_i^{vm}$ indicates the operational expenditure for booting the corresponding virtual machine for the $i$-th function whereas $opex_i^{vnf}$ indicates the operational expenditure for launching the corresponding VNF instance. Note that once a potential VNF instance is instantiated and selected, it is a deployed VNF instance whose OPEX penalty is $opex^{normal}$, which is the normal operational expenditure (e.g., normal energy consumption, etc.), if it is selected in the future.

Therefore, the OPEX penalty of chain $c$ is the sum of all VNF instances along $c$, formulated as follows:

$$pen_c^{opex} = \sum_{i=1}^{N} pen_{ij}^{opex} \quad (17)$$

With the previous definitions of QoE gain, QoS constraints penalty and OPEX penalty, we define the immediate reward associated with chain $c$ as follows:

$$r_c = gain_c^{qoe} - pen_c^{qcon} - pen_c^{opex} \quad (18)$$

For those VNF instances that participate in the construction of chain $c$ under state $s$ and action $a$, the reward $r_c$ of the chain $c$ is evenly distributed among them, formulated as follows.

$$r_{ij} = \frac{r_c}{N}, ins_{ij} \in c \quad (19)$$

## 5.4 | Action Value Function

The action value function $Q(s, a)$ is the long term discounted accumulation of immediate reward $r$. Every time a VNF instance $ins_{ij}$ is selected, immediate reward $r_{ij}$ is generated and observed as shown in equation (19), which is then used to update the current value of action value function $Q(s, ins_{ij})$ for $ins_{ij}$ as the evidence for future selections. Recall that in Section 5.2 when we defined state $s \in S$, among others, its main ingredient is the QoS/QoE status, which is non-discrete space (i.e., QoS and QoE have continuous values and change values due to resource consumption). Therefore, tabular based reinforcement learning algorithms such as Q-learning[38] are not applicable for state storage and updates. To this end, DQN (Deep Q Network)[39] is employed to fit the long term rewards $Q(s, ins_{ij})$s from a specific state $s$ and immediate rewards $r_{ij}$s.

DQN is the modification of Q-learning that uses Convolutional Neural Network (CNN) to approximate an action value function $Q(s, a)$ with a fit value $Q(s, a; \theta)$, where $\theta$ is the CNN parameter. Structurally in our framework, CNN is in charge of the state $s$ processing and $Q(s, ins_{ij}; \theta)$ generation and policy (e.g., $\epsilon$-greedy. See Section 5.5) directs $ins_{ij}$ selection. DQN have two independent networks, namely evaluation network (eval-net) and target network (target-net), which are identical in structure. However, the network parameter $\theta$ of eval-net updates after every iteration while the network parameter $\theta^-$ of target-net is frozen temporarily and updated after every $C$ iterations by $\theta$. In our framework, first of all, the loss function is defined as mean square error in the follows:

$$L(\theta) = E\left[\left(r_{ij} + \gamma \max_{ins'_{ij}} Q(s', ins'_{ij}; \theta^-) - Q(s, ins_{ij}; \theta)\right)^2\right] \quad (20)$$

Then, the gradient is derived accordingly as follows:

$$\frac{\partial L(\theta)}{\partial \theta} =$$
$$E\left[\left(r_{ij} + \gamma \max_{ins'_{ij}} Q(s', ins'_{ij}; \theta^-) - Q(s, ins_{ij}; \theta)\right) \frac{\partial Q(s, ins_{ij}; \theta)}{\partial \theta}\right] \quad (21)$$

With gradient descent and back propagation, we can acquire the optimal $Q(s, ins_{ij})$ for a specific $i$-th function.

## 5.5 | VNF Instances Selection

For a given function, the reinforcement learning agent has two strategies to select VNF instances: 1) it may prefer to choose new VNF instances that have not been executed, enhancing the perception of the network situation (i.e., exploration) and improving their probability of making optimal decisions; or 2) it may also prefer to repeatedly select currently known best VNF instances according to the current network situation and obtain the maximum known return (i.e., exploitation) with a relatively conservative approach. In fact, controllers face the Exploration-Exploitation Dilemma[40]: the risk of excessive exploration is that it is difficult to maximize rewards while excessive exploitation may lose the chance of discovering better alternatives; at the same time, the available resources may also be exhausted by excessive exploitation (e.g., some nodes might be overloaded).

To this end, it is not sufficient to conduct VNF instance selection purely based on the observed QoE/QoS which constitute the deterministic aspect (although transiently) of the network status, but certain probability models should also be adopted to capture the stochastic nature of a dynamic environment. In other words, the policy, which governs the VNF instance selection in SFC context, should take into account QoE/QoS as well as probability distributions.

The Exploration-Exploitation is often modeled as the Multi-Armed Bandit Problem (MAB)[41] in the field of reinforcement learning. MAB describes the problem that a gambler repeatedly pulls on one of the arms of a gambling machine (i.e., the bandit) for a certain amount of rewards (such as spitting a certain amount of coins). However, the distribution of each arm's reward is unknown. The goal of the MAB is to maximize the average reward after pulling arms several times. Obviously, in MAB, there is a discovery process to know the distribution of rewards for each arm (i.e., exploration) and there is also a process to maximize the average rewards after multiple arms pulling (i.e., exploitation).

Commonly seen algorithms for MAB include greedy, $\epsilon$-greedy, Softmax, UCB (Upper Confidence Bound). The greedy policy selects the action with the highest value function $Q^\pi(s,a)$ repeatedly. However, reward $r_{ij}$ (which is used to accumulate $Q^\pi(s,a)$) as we defined in SFC scenario is a non-static value, thus it is not practical to achieve long term rewards maximization using the greedy policy. $\epsilon$-greedy, however, balances exploration and exploitation by conducting the greedy policy with probability $(1-\epsilon)$ and selecting a random action with probability $\epsilon$.

$$\pi \leftarrow \begin{cases} 1 - \epsilon + \frac{\epsilon}{M_i}, & \text{if } ins_{ij} = \arg\max_{j=1}^{M_i} Q(s, ins_{ij}) \\ \frac{\epsilon}{M_i}, & \text{if } ins_{ij} \neq \arg\max_{j=1}^{M_i} Q(s, ins_{ij}) \end{cases} \quad (22)$$

Softmax adopts the Boltzmann distribution in selecting actions, which is formulated in equation (23). $\tau > 0$ is the "temperature" parameter, approximating pure exploitation when it approaches 0, while approximating pure exploration when it approaches 1. $\tau$ offers the possibility to balance between exploration and exploitation. Meanwhile, $\tau$ can be decremented to reduce exploration in a later phase when the convergence is reached.

$$\pi \leftarrow \frac{e^{\frac{Q(s,ins_{ik})}{\tau}}}{\sum_{j=1}^{M_i} e^{\frac{Q(s,ins_{ij})}{\tau}}} \quad (23)$$

UCB is another common algorithm for policy enforcement. UCB takes the form of (mean + upper confidence bound) as shown in equation (24), which is also the reason for its naming. According to the large number theorem, the mean can be replaced by the arithmetic average as shown in the first part on the right; the upper confidence bound is given by the Chernoff-Hoeffding inequality as shown in the second part on the right. $count_{ij}$ represents the number of times $ins_{ij}$ was selected and $count$ represents the total number of SFC requests that have been solved. One of the advantages offered by UCB is that the workload is adaptively balanced between VNF instances $ins_{ij}$s of the same VNF type $t_i$. The more times a VNF instance $ins_{ij}$ is selected, the greater value $count_{ij}$ is, leading to the decrease of upper confidence bound. For those VNF instances with similar $Q(s_i, a_{ij})$, the smaller value $count_{ij}$ has, the greater chance it is selected, thus the workload is adaptively balanced.

$$\pi \leftarrow \max_{j=1}^{M_i} \left( Q(s, ins_{ij}) + \sqrt{\frac{2\log(count)}{count_{ij}}} \right) \quad (24)$$

## 5.6 | QoS/QoE-aware SFC Algorithm

Based on the previous modeling, we give the DQN_Q2_SFC training algorithm in Algorithm 1. $Q$ is the $M_i$-dimensional vector of $Q(s, ins_{ij})$s for the $i$-th function. $EPI\_COUNT$ is the number of training episodes. $REC\_COUNT$ is the number of SFC requests used for training. Note that in line 9, the selection of VNF instances is restricted in that $ins_{ij}$ must be connected to the current instance so as to adapt to the topological dynamics (e.g., instance shutdown, etc.). Note also that, in our real implementation, we use the $\epsilon$-greedy (i.e., equation (22)), therefore, we can balance between exploration and exploitation by tuning hyper parameter $\epsilon$.

---

**Algorithm 1** DQN_Q2_SFC
---
1: initialize replay memory $D$ to capacity $N$
2: initialize action value function $Q$ with random weights $\theta$
3: initialize target action value function $\hat{Q}$ with weights $\theta^-$
4: **for** $episode = 1..EPI\_COUNT$ **do**
5:     reset environment
6:     **for** $sfc\_req = 1..REQ\_COUNT$ **do**
7:         initialize chain $c$ and observe initial observation $s$
8:         **for** $i = 1..N$ **do**
9:             select a connected instance $ins_{ij}$ by eq. (22), (23) or (24)
10:            observe $s$ by QoS over LLDP, etc. and observe $r_{ij}$ by eq. (19)
11:            store transition $(s, ins_{ij}, r_{ij}, s')$ in $D$
12:            if $s$ is terminal state, break
13:            $s = s'$
14:         **end for**
15:         sample minibatch of transitions $(s, ins_{ij}, r_{ij}, s')$ from $D$
16:         every $C$ iterations, reset $\hat{Q} = Q$
17:         update $Q$ by gradient descent (eq. (20), (21))
18:     **end for**
19: **end for**

## 6 | EXPERIMENTS

The experiment environment is as follows: Ubuntu 14.04 Server with 64 GB memoery, 40 logical CPUs with 1200 MHz, 2 Tesla GPUs (only one used during experiments), and TensorFlow 1.0.0 (compiled on GPU).

### 6.1 | Experiments on QoS over LLDP

The QoS over LLDP is deployed in Mininet and Floodligth 1.3 environment. It is tested in a topology (see Figure 4 ) where a video streaming application is deployed. Host h1 is the video server, hosting a video (about 500 MB in size and 15 min in length) and h3 is the client, streaming the video from h1 using Firefox browser. During the streaming, QoS status is constantly changing in bandwidth, delay, etc., due to resource consumption. In order to evaluate the impact on network traffic caused by QoS over LLDP, we capture traffic using Wireshark in two scenarios (video streaming v.s. no video streaming) with the above topology. The evaluation duration is 15 min. The evaluation results are shown in Table 2 .

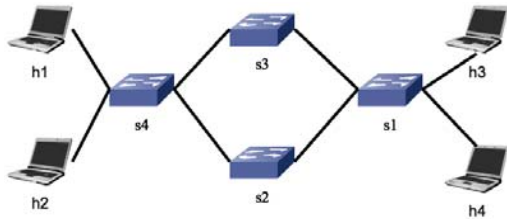

**FIGURE 4** The Topology for QoS over LLDP Experiment

We first analyze the "no video streaming" scenario. As we stated above, QoS over LLDP causes extra network overhead since it contains several QoS TLV bytess. However, the percentages of QoS over LLDP (6.56%) is just slightly greater than pure LLDP (5.27%) by bytes, meaning that QoS over LLDP does not deteriorate the network traffic performance. For the "video streaming" scenario, both QoS over LLDP and pure LLDP take almost the same percentage, 0.021% and 0.015% of the total traffic by bytes, respectively and by packets, 0.59% and 0.58% of the total traffic by packets, respectively. This indicates that QoS over LLDP works in a piggyback fashion with very minor traffic overhead to achieve QoS information collecting. The experiment results indicate that QoS over LLDP is an applicable approach for QoS information delivering in an SDN environment.

### 6.2 | Experiments on DQN-based QoS/QoE-aware Service Function Chaining

Our DQN-based QoS/QoE-aware SFC algorithm is tested by TensorFlow simulation in this section. We compare our algorithm with violent search, which guarantees best service function chain with the highest QoE, and random search, which gives a functionally feasible chain with minimal response time. The experiment topology contains 10 VNF types where each type contains 10 VNF instances whose QoS status are generated randomly when the network topology is initialized. The purpose is to compare their performance in QoE provisioning, QoS constraining, response time, etc. 250 episodes are conducted in the comparison, where one episode includes 100 SFC requests. The experiment results are shown in Figure 5 , 6 and Table 3 .

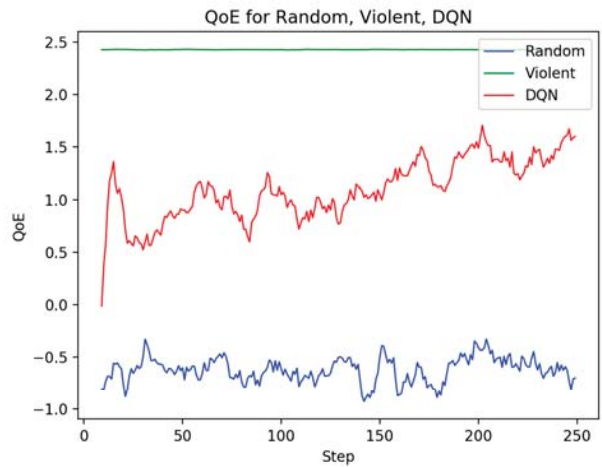

**FIGURE 5** QoE Comparison between Random, Violent and DQN-based

We can see from Figure 5 that our DQN-based algorithm orchestrates service function chains with QoE between that of random search and violent search. Violent search ensures best QoE which DQN gradually approaches. This indicates that DQN exhibits a strong learning ability to approximate the best QoE after training. Random search gives only functionally feasible chains without QoE provisioning, thus it is often penalized in terms of QoE, shown in Figure 5 . Meanwhile, with regard to QoS constraining, the DQN-based algorithm respect QoS constraints with overwhelming probability. The reason why there are still cases where the DQN-based algorithm violates QoS constraints is that the QoS constraints vector is modeled as penalty (i.e., a scalar) against reward, leading to precision and dimension losses that eventually cause violations. Therefore, our algorithm can be somehow seen as heuristics to violent search, in exchange for agile orchestration that still possesses a high QoE. Note that, however, the probability that DQN-based algorithm violates QoS constraints gradually lowers down, which again exhibits its strong learning ability. In this regard, DQN-based algorithm balances QoE provisioning and QoS constraining. With regard to response time (Table 3 ), violent search delivers response time orders of magnitude slower than that of DQN, which is not acceptable in practical applications, especially for time-critical applications. DQN is quick to response in that it gives almost constant time complexity after agent training.

**TABLE 2** QoS over LLDP v.s. LLDP in different scenarios, duration about 15 min.

| Scenario | Scheme | Total Packets | LLDP Packets | Total Bytes | LLDP Bytes |
| --- | --- | --- | --- | --- | --- |
| No Video Streaming | Pure LLDP | 20173 | 2567 (12.72%) | 6526386 | 344222 (5.27%) |
| | QoS over LLDP | 21889 | 2563 (11.71%) | 6738255 | 441760 (6.56%) |
| Video Streaming | Pure LLDP | 440348 | 2550 (0.58%) | 2208747286 | 339866 (0.015%) |
| | QoS over LLDP | 437795 | 2580 (0.59%) | 2086069045 | 443204 (0.021%) |

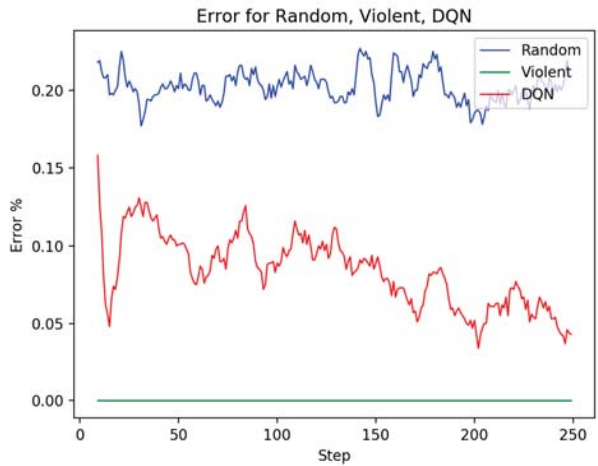

**FIGURE 6** QoS Comparison between Random, Violent and DQN-based

**TABLE 3** Response Time Comparison between Random, Violent and DQN-based.

| Algorithms | Response Time |
| --- | --- |
| Random | 0.00019 |
| Violent | $> 15 min$ |
| DQN | 0.004 |

## 7 | CONCLUSIONS

In this paper, we propose a reinforcement learning (DQN, to be exact) based QoS/QoE service function chaining framework for SDN/NFV-enabled 5G slices. It features two aspects, i.e., 1) the lightweight QoS over LLDP scheme to bring QoS awareness and 2) the DQN-based SFC algorithm that synthetically takes into account QoS and QoE as key ingredients to formulate rewards. The experiments show that it is applicable in service function chaining in 5G core network slices in dynamic QoS environments.

In our current work, we focus on the QoS/QoE-aware SFC in single 5G slices. In a service outsourcing scenario, a service function chain might involve trans-slice network functions, which are likely to introduce hierarchical orchestration design. We consider this as a future work direction. Meanwhile, our current work focuses on service function chaining in 5G core network slices. Our next step will also address to coordinate MEC in 5G edge networks with SFC in 5G core networks to bring end-to-end QoE-satisfactory services is also our next step work.